      [1999/12/01 v1.4c Il Nuovo Cimento]
\documentclass{cimento}
\usepackage{epsfig}
\title{Testing fundamental physics with high-energy cosmic rays}
\author{John Ellis\from{ins:TH}\thanks{Talk at the Chacaltaya Meeting On
Cosmic Ray Physics, 23-27 July 2000, La Paz, Bolivia}}
\instlist{
  \inst{ins:TH} 
Theoretical Physics Division, CERN, CH-1211 Geneva, Switzerland
               }

\PACSes{\PACSit{96.40}{Cosmic rays}
\PACSit{01.30.Cc}{Conference proceedings}}

\def\beq{\begin{equation}}
\def\eeq{\end{equation}}
\def\bea{\begin{eqnarray}}
\def\beaa{\begin{eqnarray*}}
\def\eea{\end{eqnarray}}
\def\eeaa{\end{eqnarray*}}
\def\bq{\begin{quote}}
\def\eq{\end{quote}}
\def\gappeq{\mathrel{\rlap {\raise.5ex\hbox{$>$}}
{\lower.5ex\hbox{$\sim$}}}}

\def\lappeq{\mathrel{\rlap{\raise.5ex\hbox{$<$}}
{\lower.5ex\hbox{$\sim$}}}}

\begin{document}
\maketitle

\begin{abstract}

Cosmic rays may provide opportunities for probing fundamental physics. 
For example, ultra-high-energy cosmic rays might originate from the decays
of metastable heavy particles, and astrophysical $\gamma$ rays can be used
to test models of quantum gravity. Both scenarios offer ways to avoid the
GZK cut-off. \\
\begin{center}
CERN-TH/2000-314 ~~~~~~ - ~~~~~~ astro-ph/y0010474
\end{center}

\end{abstract}

\subsection{Introduction}

\begin{figure}[htbp]
\begin{center}  
\mbox{\epsfig{file=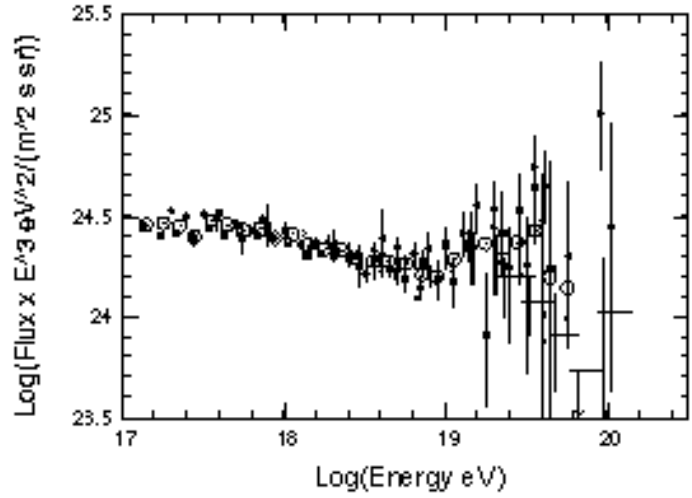,height=7cm}}
\end{center}
\caption[.]{\label{fig:UHECR}\it
The ultra-high-energy tail of the cosmic-ray spectrum, which does not
turn down as expected on the basis of the GZK cut-off~\cite{GZK}.
}
\end{figure}

In this lecture I discuss two cosmic-ray topics where fundamental physics
may be testable: the ultra-high-energy cosmic rays that apparently evade
the GZK cut-off may probe the decays of
superheavy particles, and astrophysical sources of $\gamma$ rays may probe
quantum gravity.  At first sight, there is no obvious relations between
the two subjects.  However, as we see at the end, quantum-gravity effects
might provide another mechanism for evading the GZK cut-off: see, for both
ultra-high-energy cosmic rays (see Fig.~\ref{fig:UHECR}) and astrophysical
$\gamma$ rays. 

\subsection{A
Top-Down Decay
model for
Ultra-High-Energy Cosmic Rays}

\begin{figure}[htbp]
\begin{center}
\mbox{\epsfig{file=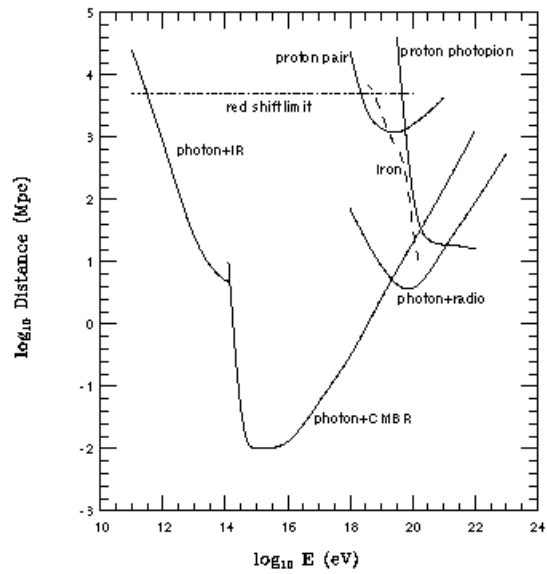,height=8cm}}
\end{center}
\caption[.]{\label{fig:cutoffs}\it
The cut-offs expected for high-energy protons~\cite{GZK}, Iron
nuclei and photons, due to photo-absorption processes and
$e^+ e^-$ pair production, respectively~\cite{LS}.
}
\end{figure}

\begin{figure}[htbp]
\begin{center}
\mbox{\epsfig{file=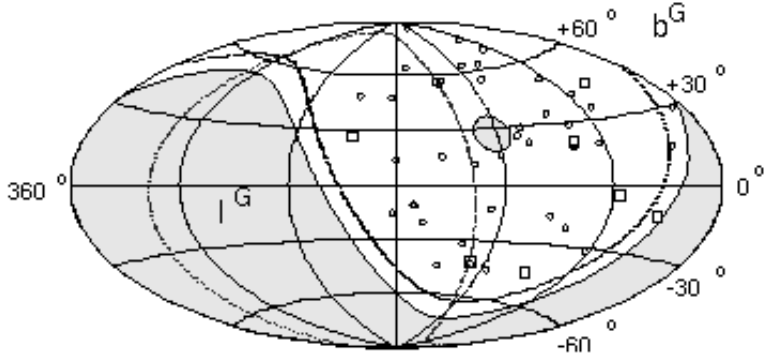,height=5cm}}
\end{center}
\caption[.]{\label{fig:directions}\it
The arrival directions of ultra-high-energy cosmic rays seen by the AGASA
array: the shaded regions are invisible to this experiment~\cite{LS}.
}
\end{figure}

As is well known, one expects a suppression of ultra-high-energy (UHE)
protons above $E \sim 5 \times 10^{19}$ eV, due to absorption by cosmic
microwave background photons: $p + \gamma_{CMBR} \to \Delta^+$~\cite{GZK}:
this and
analogous cut-offs for $Fe$ and $\gamma$'s are  seen in
Fig.~\ref{fig:cutoffs}~\cite{LS}. 
However, no such effect is seen (Fig.~\ref{fig:UHECR}) in the
data~\cite{noGZK}, suggesting that
these cosmic
rays must originate from nearby: $d \lappeq 100$ Mpc for $E \sim 10^{20}$
eV.  In this case, unless magnetic field effects are unexpectedly
strong,
one would expect the UHE cosmic rays to point back to astrophysical
sources.  However, as seen in Fig.~\ref{fig:directions}, no clear evidence
for any discrete sources has yet emerged, although there are some
candidate doublets and triplets whose statistical significance is not yet
overwhelming~\cite{LS} - see, however~\cite{BF}.

Under these
circumstances, it
is natural to explore
possible origins in
new physics.  In particular, the decays of supermassive, metastable dark
matter particles~\cite{Gelmini} clustered in our galactic halo 
might lead to ultra-high-energy cosmic rays that would evade the GZK
cut-off~\cite{evade}, have no discrete sources, and yield an approximately
isotropic distribution. 
One of the issues in such a top-down model is why the supermassive
particle should be metastable~\cite{ELN}. 

An example of analogous metastability may be provided by the proton. The
Standard Model symmetries do not admit any renormalizable interaction
violating baryon number, but this is an accidental consequence of other
symmetries of the theory: higher-dimension $\Delta
B \not= 0$ interactions are permitted. In traditional GUTs, the first such
interaction has dimension 6, and gives an amplitude scaled by $1/M^2$,
where $M$ is the GUT mass scale, leading to a proton lifetime $\tau \sim
M^4/m^5_p$.  This may exceed $10^{33}y$ if $M \gappeq 10^{15}$ GeV. 

In many models, relic particles decay through higher-dimensional operators
$\sim 1/M^n$, in which case the lifetime $\tau \sim
M^{2n}/m^{2n+1}_{Relic}$.  This may comfortably exceed $10^{10} y$ if $M$
and $n$ are large enough, e.g., $M \sim 10^{17}$ GeV and $n \geq 9$ are
sufficient if $m_{relic} \sim 10^{12}$ GeV. 

Constraints on such metastable relics come from light-element abundances,
the cosmic microwave background and the high-energy astrophysical $\nu$
flux~\cite{Gelmini}.  An abundance of relic particles weighing $10^{12}$
GeV sufficient
to yield $\Omega_{Relic}h^2 \sim 1$ is possible if $\tau \sim 10^{16} y$. 

Is it at all possible or plausible that a superheavy relic might have
$\Omega_{Relic}h^2 \sim 1$?  With the standard mechanism of freeze-out
following thermal equilibrium, one would expect that $m_{Relic} \lappeq 1$
TeV in order to obtain an interesting relic density.  However, it has
recently been realized that this upper limit may be avoided by non-thermal
and gravitational production mechanisms around and after
inflation~\cite{CKR}.
Depending on the details of the model, $\Omega_{Relic}h^2 \sim 1$ may be
possible for $10^8 \; \rm{GeV} \lappeq m_{Relic} \lappeq 10^{18}$ GeV. 

We have explored possible candidates for such a metastable superheavy
relic in string/M theory~\cite{BEN}. These contain Kaluza-Klein states
(`hexons')
that acquire masses when $10 \to 4$ or $11 \to 5$ dimensions, but these
are not expected to be metastable, and may be too heavy.  In $M$ theory,
more Kaluza-Klein states appear when $5 \to 4$ dimensions (`pentons'), but
these are also expected to be very unstable.  The last candidates may be
bound states from a hidden sector of string/M theory
(`cryptons')~\cite{ELN}. Their masses are determined by the
non-perturbative dynamics of this hidden sector, and they
may well have masses in the range $m_{Relic} \sim 10^{12}$ to $10^{13}$
GeV of interest for ultra-high-energy cosmic rays.  For example, in the
flipped SU(5) model derived from string, the hidden-sector gauge group is
$SU(4) \times SO(10)$, and some of the states bound by the former factor
(`tetrons') decay via higher-dimensional operators, plausibly weighing
$\sim 10^{12}$ GeV and with lifetimes $\gappeq 10^{15} y$~\cite{BEN}. 

\begin{figure}[htbp]
\begin{center}
\mbox{\epsfig{file=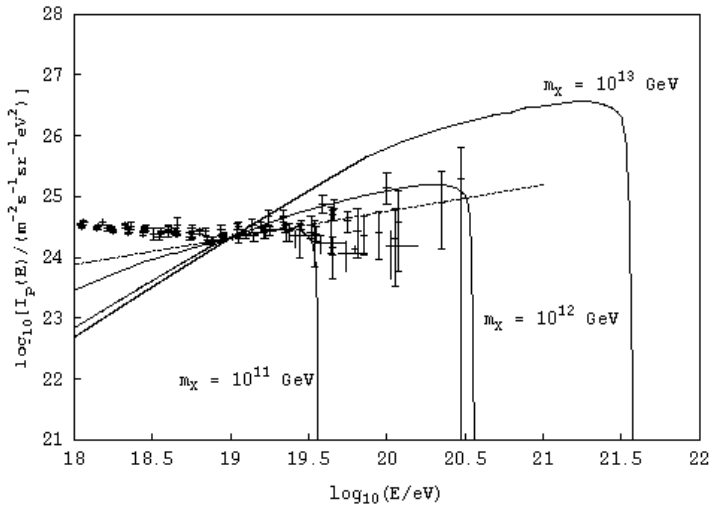,height=7cm}}
\end{center}
\caption[.]{\label{fig:BS}\it
The observed spectrum of ultra-high-energy cosmic rays compared with
a calculation of crypton decays, for various different choices of the
crypton mass~\cite{BS}.
}
\end{figure}

The hadronization of quarks produced in crypton decay has been modelled
both with~\cite{Kach} and without supersymmetry~\cite{BS}, and the
spectrum appears compatible
with the few ultra-high-energy cosmic rays observed, as seen in
Fig.~\ref{fig:BS}. The
Auger~\cite{Auger} and EUSO~\cite{Scarsi}
projects offer the best prospects for distinguishing cryptons from
alternative explanations of the existing data~\cite{CLAF}. 

\subsection{Space-Time Foam}

We know that space-time is quite flat on large distance scales.  For
example, we learnt Euclidean geometry at school, not the Riemannian
geometry of curved space, and cosmological microwave background
experiments indicate that the Universe is flat on a scale of $10^{10}$
light years. However, in any quantum theory of gravity one expects large
fluctuations in the fabric of space-time at the Planck scale: fluctuations
in the energy $\Delta E \sim m_P \sim 10^{19}$ GeV, accompanied by
topology changes $\Delta \chi \sim 1$, over distance scales $\Delta x \sim
l_P \sim 10^{-33}$ cm, lasting for times $\Delta t \sim t_P \sim
10^{-43}$s~\cite{Rovelli}. 

\begin{figure}[htbp]
\begin{center}
\mbox{\epsfig{file=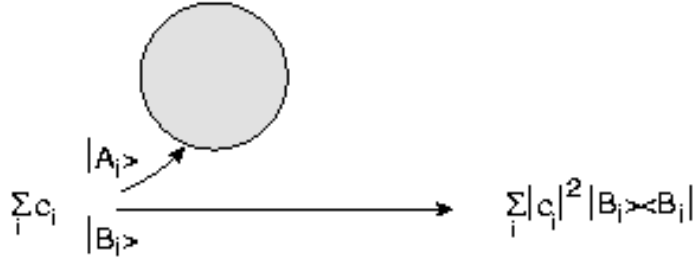,height=4cm}}
\end{center}
\caption[.]{\label{fig:EMN1}\it
A process which (might) lead to quantum decoherence in the
neighbourhood of a (microscopic) black hole: information about the phase
of component $|A>$ of the wave function is lost across the (microscopic)
horizon~\cite{Hawking,EHNS}.
}
\end{figure}

Are there any observable consequences of such microscopic
quantum-gravitational fluctuations~\cite{Hawking}?  Does loss of
information occur across
microscopic event horizons, modifying conventional quantum mechanics so as
to allow pure quantum-mechanical states to evolve into mixed
ones~\cite{EHNS}, as suggested by Fig.~\ref{fig:EMN1}?
And,
more relevant to this meeting, as a particle passes by, does its
gravitational effect make
the vacuum react to its energy, and does this recoil of the vacuum reduce
the effective velocity of the particle~\cite{AEMN}:
\beq 
c(E) \simeq
c_0 (1 - E/m_{QG} + \dots ) \;?  
\label{twoone}  
\eeq 
Here, $c_o$ is the
`classical' (low-energy) velocity of light and $m_{QG}$ is some high mass
scale that might be ${\cal O}(m_P)$. 

\begin{figure}[htbp]
\begin{center}
\mbox{\epsfig{file=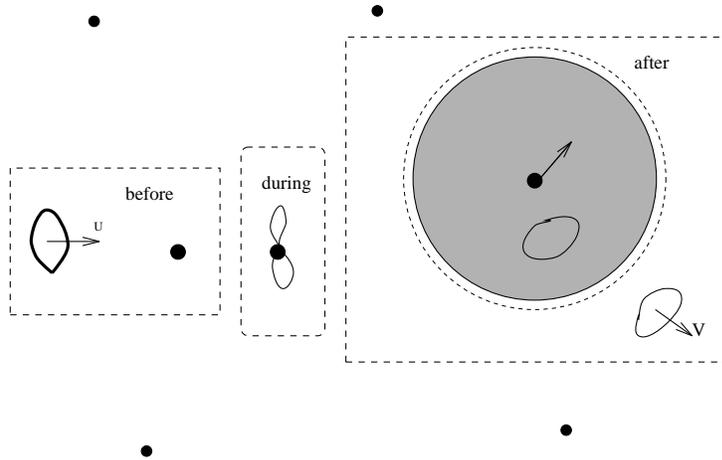,height=6cm}}
\end{center}
\caption[.]{\label{fig:EMN2}\it
Sketch of a (closed-string) particle state hitting a ($D$-brane)
defect in space-time. After impact, the defect is excited and an
expanding horizon is formed. Particles
may be trapped inside, losing information, and the propagation of
the incident particle may be slowed~\cite{recoil}.
}
\end{figure}

In order to address such questions, one must formulate a model of
space-time foam.  We imagine that it contains virtual topological
`defects', ${\cal O}(1)$ per Planck-size four-volume, appearing and
disappearing as quantum fluctuations, perhaps like instantons in QCD.  We
model these defects as solitonic $p$-dimensional `lumps' of string called
$D$ particles, or more generally $Dp$ branes~\cite{Dmodel}, as seen
in Fig.~\ref{fig:EMN2}. The
formal
technology of
string theory can be used to describe the interaction of an energetic
particle hitting such a `lump'.  Among the effects of such a collision are
a modified (reduced) velocity for the energetic particle, recoil motion of
the struck defect and, at the quantum level, excitation of the defect to a
higher state~\cite{recoil}. 

We argue that
the recoil of the defect does indeed modify the background metric by an
amount
\beq
h_{o_i} \propto \bar{U}_i \, \Theta(t)
\label{twotwo}
\eeq
at large times $t$, where $\bar{U}_i$ is the defect's recoil velocity, and
the collision is assumed to have taken
place at time $t = 0$.  One may consider the propagation of a
photon~\cite{EMN} or other (almost)
massless particle, such as a neutrino~\cite{EMNV}, by computing the null
geodesic~\cite{EMNnull}:
\beq
c^2_0 (dt)^2 = (dx)^2 + 2h_{o_i} dt dx_i \, ,
\label{twothree}
\eeq
using (\ref{twotwo}) for $h_{o_i}$.  For symmetry reasons, the
effective $\bar{U}_i$ must be in the direction of the particle motion.
Labelling its magnitude by $\bar{U}$, one
finds
\beq
c_0 \, \frac{dt}{dx} \simeq \bar{U} + \sqrt{1 + \bar{U}^2} \simeq 1 + \bar{U} + \cdots
\label{twofour}
\eeq
implying a {\it reduced} velocity:
\beq
c(E) = s\frac{dx}{dt} = c_0 \, (1 - \bar{U} + \cdots ).
\label{twofive}
\eeq
As a rough order-of-magnitude estimate, one may guess that the recoil momentum $k \sim E$, so
that the recoil velocity $\bar{U} \sim k/M$, where $M$ is the effective mass of the defect,
that one could expect to be ${\cal O}(m_P)$.  Thus one finds $U \sim E/M$
and hence
\beq
c(E) \simeq c(1 - E/M + \cdots ),
\label{twosix}
\eeq
similar to that suggested for photons~\footnote{A different suggestion 
is made in~\cite{Mex}.}.

A similar result can be derived from Maxwell's equations~\cite{Mitsou}:
\bea
\underline{\nabla} \cdot \underline{B} = 0, &
\underline{\nabla} \times \underline{H} = 
\frac{1}{c_0} \, \frac{\partial}{\partial t} \, \underline{D} = 0 \nonumber \\
\underline{\nabla} \cdot \underline{D} = 0, &
\underline{\nabla} \times \underline{E} = 
- \frac{1}{c_0} \, \frac{\partial}{\partial t} \, \underline{B} = 0
\label{twoseven}
\eea
in empty space, where
\beq
\underline{D} = 
\frac{\underline{E}}{\sqrt{h}} + \underline{H} \times \underline{G}, \;
\underline{B} = \frac{\underline{H}}{\sqrt{h}} + \underline{G} \times \underline{E}
\label{twoeight}
\eeq
with 
\beq
G_{oo} = -h, \; 
\frac{G_{oi}}{G_{oo}} = -G_i.
\label{twonine}
\eeq
Using (\ref{twotwo}), it is easy to find the following wave equations, to
leading order:
\beq
\left(
\frac{1}{c^2_0} - \underline{\nabla}^2 - 2 (\underline{\bar{U}} \cdot \underline{\nabla}) \;
\frac{1}{c_0} \; \frac{\partial}{\partial t} \right )
(\underline{B}, \underline{E}) = 0
\label{twoten}
\eeq
If one looks for a plane-wave solution:
\beq
E_x = E_z = 0, E_y = E_o \, e^{i(kx-wt)},
B_x = B_y = 0, B_z = B_o \, e^{i(kx-wt)},
\label{twoeleven}
\eeq
one finds the modified dispersion relation
\beq
k^2 - w^2 - 2 \bar{U} kw = 0.
\label{twotwelve}
\eeq
This leads, as before, to `subluminal' propagation at a velocity $c(E)
\simeq c_0 (1-\bar{U} +
\cdots )$.

There are other ways of seeing how such a non-trivial refractive index
{\it in vacuo} might arise.  For example, a particle hitting such a defect
creates an expanding horizon. It is possible for particles to be trapped
inside this horizon, providing a possible mechanism for loss of
information~\cite{EHNS}.  However, for our present purposes, the relevant
observation
is that the particle slows down as it passes through the horizon, 
as seen in Fig.~\ref{fig:EMN2}, in line
with the estimate (\ref{twotwo})~\cite{horizon}. 

A similar effect appears if one considers our three-dimensional space as a
membrane in a higher-dimensional space-time.  This may fluctuate
spontaneously by emitting or absorbing string states propagating through
the `bulk' extra dimensions. The passage of an energetic photon or other
particle along our three-brane will in general modify these interactions
with `bulk' degrees of freedom, distorting the three-brane and modifying
the propagation of the photon~\cite{higherD}.  As a result, the photon
experiences a
stochastic time delay: 
\beq
\delta t \simeq g_s \, \frac{L.E}{M_s}
\label{twothirteen}
\eeq
where $g_s$ is the string coupling and $M_s$ the string scale.  This
effect is below present experimental upper limits in conventional strings
with $g_s ={\cal 0}(1)$ and $M_s \sim m_P$. However, it might be
problematically large in some low-scale string models with $M_s << m_P$,
unless also $g_s << 1$. 

There are other approaches to the modelling of space-time foam.  For
example, it has been proposed that quantum gravity be treated as a
`thermal bath' that provides a decohering medium~\cite{thermal}.  Another
interesting
approach is that of loop gravity, which yields a cellular structure in
space-time reminiscent of spin networks~\cite{loop}.  Its vacuum may be
characterized
as a `weave state' $|w>$ with the property that
\beq
<w|G_{\mu \nu}|w> = \eta_{\mu \nu} + {\cal O}(E L_w)
\label{twofourteen}
\eeq
where $L_w$ is the `weave-length' at which quantum gravity appears.  Both
the above approaches lead to a breaking of Lorentz invariance.  The vacuum
may have other non-trivial optical and thermal properties, leading, e.g.,
to birefringence~\cite{GP}.  One may also expect light-cone fluctuations,
as found
in our own approach to space-time foam. 

Quantum-gravity effects can be distinguished from those of a conventional
plasma by their different energy dependences.  The former should increase
with energy, while the latter should decrease at high energies.  Consider,
for example, photon propagation in a thermal electromagnetic plasma at
temperature $T$: 
\beq
E^2 = q^2 + \pi_T : v \equiv \frac{\partial E}{\partial q}.
\eeq
In the limit $T << q, qT >> m^2_e$, one finds
\beq
v \simeq c \left [ 1 - \frac{\alpha^2}{6} \left( \frac{T}{q} \right)^2 \;
\ell n^2 (qT/m^2_e ) + \cdots \right ]
\eeq
so that~\cite{plasma1}
\beq
v(E) \simeq c \left [ 1 - {\cal O} \left( \frac{1}{E^2} \, \ell n E \right ) \right ]
\eeq
Even smaller effects are expected if only the background photons are thermalized, as in the
Universe today.  One finds~\cite{plasma2}
\beq
v \simeq c(1-\gamma ) : \gamma = \frac{44 \pi^2}{2025} \, \alpha^2 \left (
\frac{T}{m_e} \right )^4 ,
\eeq
i.e., a constant reduction in velocity that is unobservably small in the present Universe with
$T \simeq 2.7^{\circ} K$.

\subsection{Astrophysical Probes of the Velocity of Light}

Astrophysical sources, many of which are at cosmological distances, offer
some of the best prospects for probing the possible energy dependence of
the velocity of light:  $c(E) \simeq c_0 (1 - E/M + \cdots )$ and a
possible stochastic spread in velocities of photons of given energy:
$\delta c \sim c_0 E/\Lambda + \cdots $~\cite{AEMNS}. These effects would
yield at a time delay (or spread): 
\beq
\delta t \simeq \frac{L}{c_0} \, \frac{E}{M \,{\rm or} \, \Lambda}
\eeq
where $L$ is the distance of propagation.  The figure of merit for probing
$M$ (or $\Lambda$)  is clearly $L.E/\delta t$, i.e., one wants distant,
high-energy sources with short intrinsic time-scales.  Examples of
interesting sources are pulsars ($L \sim 10^4$ light years, $E \sim 1$
GeV, $\delta t \sim 300$ s), active galactic nuclei (AGNs) ($L \sim 100$
Mpc, $E \sim 2$ TeV, $\delta t \sim 300$ s) and gamma-ray bursters (GRBs)
($L \sim 10^{10}$ light-years, $E \to$ TeV?, $\delta t \to 10$ ms),
providing prospective sensitivities to $M$ (or $\Lambda$) in the range
$10^{15}\, {\rm to} \, 10^{18}$ GeV~\cite{AEMNS}. 

\begin{figure}[htbp]
\begin{center}
\mbox{\epsfig{file=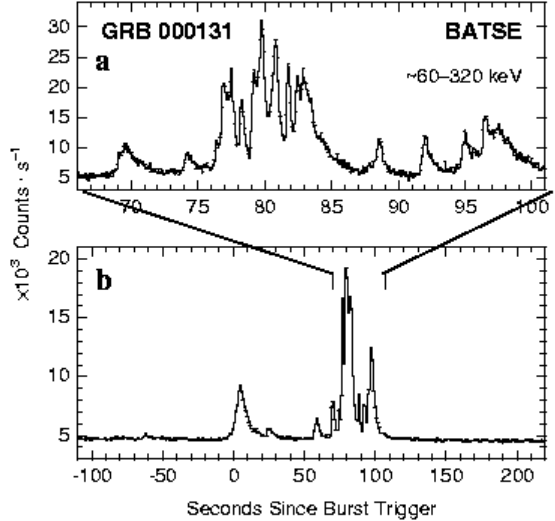,height=7cm}}
\end{center}
\caption[.]{\label{fig:GRB}\it
A spectacular GRB with a large redshift $z = 4.5$, whose pulse
exhibits many microbursts~\cite{fourpointfive}.
}
\end{figure}

Some of the best prospects may be offered by GRBs~\cite{AEMNS,Mitsou}.
One is visible per day
on average, throughout the Universe, and BATSE has recorded almost 3000
GRB triggers. Their durations range between seconds and hundreds of
seconds, and some exhibit microburst structures on the millisecond scale:
a spectacular recent example is shown in Fig.~\ref{fig:GRB}.
Several have now been seen at other wavelengths (energies) including radio
(afterglows), optical (in flagrante and afterglows), X-ray (afterglows)
and possibly TeV $\gamma$ rays (in flagrante).  
The afterglow observations
have confirmed that there is a substantial population (at least of
multi-second GRBs) with high redshifts $z = {\cal O}(1)$, ideal for our
purpose! 

There are several models for GRBs on the market, including mergers of
neutron stars and/or black holes, anisotropic supernovae that squirt
in particular directions, and
hypernovae (collapses of massive stars)~\cite{GRBs}. The internal engine
is of
secondary importance to us, as is its possible anisotropy.  What is
important to us is that all models agree on the formation of a highly
relativistic ($\gamma = {\cal O}(100)$) optically thin plasma that
exhibits
stochastic fluctuations on short time scales, presumably because of
internal shocks. 

\begin{figure}[htbp]
\begin{center}
\mbox{\epsfig{file=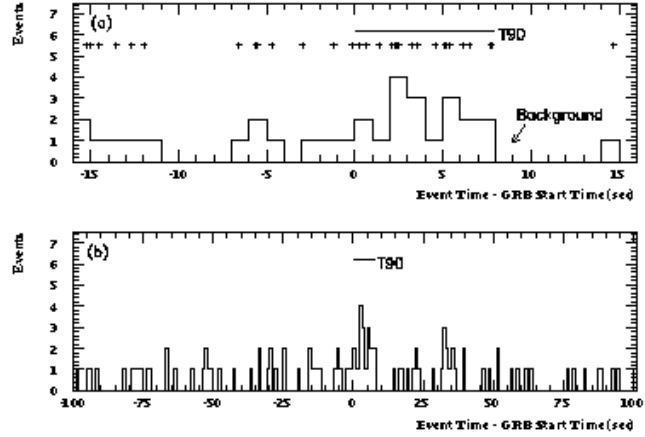,height=6cm}}
\end{center}
\caption[.]{\label{fig:Milagrito}\it
The Milagrito time series at the time of GRB 970417a, indicating a
possible concidence with TeV photons~\cite{Milagrito}.
}
\end{figure}

We would like to probe the simultaneity of these pulses in different
energy bands.  For example, BATSE observed GRBs in four bands ($25$ to
$50$ keV, $50$ to $100$ keV, $100$ to $300$ keV and $> 300$ keV) and OSSE
(also aboard the CGRO) observed at energies $> 2$ MeV.  EGRET has reported
some multi-GeV photons, for example with $E \sim 30$~GeV during
the first $200$ ms of GRB 930131~\cite{EGRET}.  Most exciting is the
report of a
possible signal in TeV photons coincident with GRB
970417a~\cite{Milagrito}, shown in Fig.~\ref{fig:Milagrito}. 
If one
assumes that the source was at $z \sim 0.1$, that $\delta t \sim 10$s and
$E_{\gamma} \sim 1$ TeV, one finds a sensitivity to $M \gappeq 10^{18}$
GeV.  Unfortunately, the statistical significance of this first Milagrito
event was not overwhelming, its redshift was not measured, and no more
such coincidences have been reported so far. 

Several pioneering analyses have been made, but each is subject to some
question. The analysis of~\cite{Schaefer}
uses GRBs whose redshift was not measured, and whose distances are
therefore unknown. The analysis of a flare of Mkn 421 in~\cite{Biller}
is sensitive to $M \sim 4 \times 10^{16}$~GeV, but the detection of
the flare is not secure, having a 5\% probability of being spurious,
as mentioned by the authors themselves.
On the other hand, the $\gamma$-ray signal from the Crab pulsar used
by~\cite{Kaaret}
is statistically secure, but there is a known time difference between the
$\gamma$ and radio pulses, and this is comparable to variable dispersion 
effects in the interstellar medium. A joint statistical analysis of
signals from several pulsars would be needed to disentangle possible
source and medium effects.

\begin{figure}[htbp]
\begin{center}
\mbox{\epsfig{file=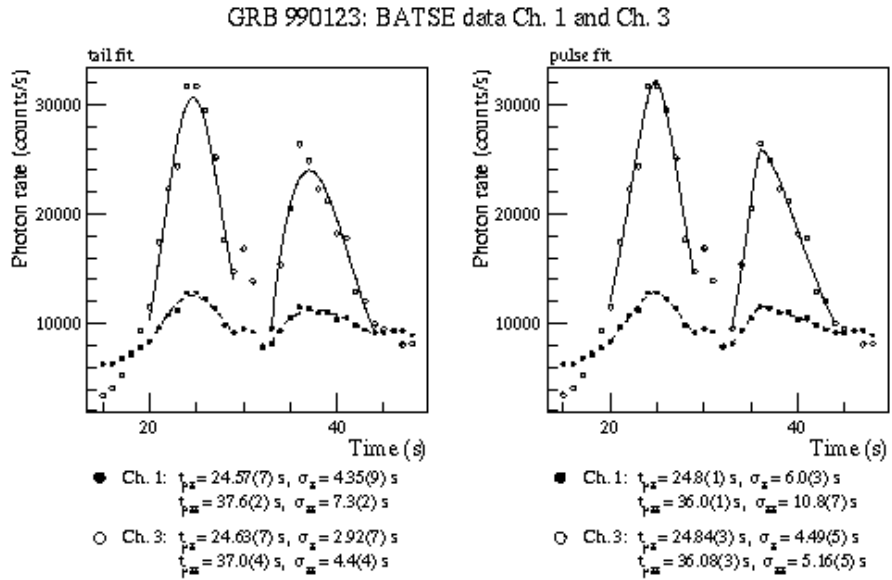,height=8cm}}
\end{center}
\caption[.]{\label{fig:Mitsou1}\it
Fits to the pulses of GRB 990123 in different BATSE energy channels,
using two different fitting functions~\cite{Mitsou}.
}
\end{figure}

\begin{figure}[htbp]
\begin{center}
\mbox{\epsfig{file=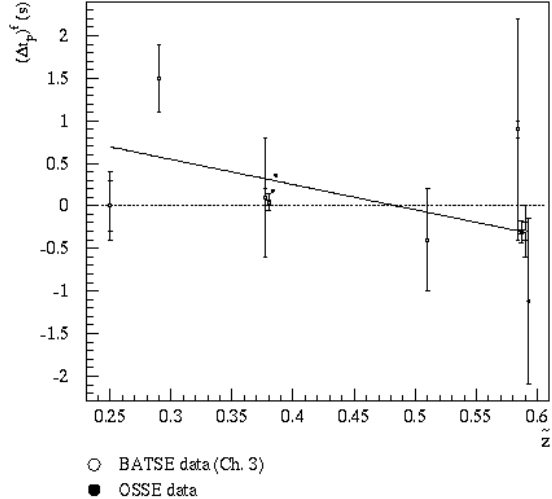,height=7cm}}
\end{center}
\caption[.]{\label{fig:Mitsou2}\it
A fit to the time-lags extracted from the pulses of GRBs with
measured redshifts. No significant dependence on the redshift
was found~\cite{Mitsou}.
}
\end{figure}

We made a systematic analysis~\cite{Mitsou} of all the GRBs with measured
cosmological
redshifts, analyzing BATSE and OSSE data and comparing arrival times in
the highest- and lowest-energy channels. We tried
several different fitting functions for the peaks observed in the
different channels, comparing their positions and widths, as seen in
Fig.~\ref{fig:Mitsou1}.  We
then looked
for a possible correlation between the time lags (spreads) and the light
travel distances, which increase with redshift, as seen in
Fig.~\ref{fig:Mitsou2}.
A real propagation
effect
should exhibit such a correlation, but a source and/or selection effect
need not.  
The data were consistent with the absence of any such
correlation, allowing us to conclude that, parametrizing a subliminal
velocity $v (E) \simeq c_o (1-E/M)$: 
\beq
M > 10^{15}\,{\rm GeV}
\eeq
and parametrizing a stochastic spread in velocities at fixed energy 
$\delta v (E) \simeq C_o \cdot E/\Lambda$:
\beq
\Lambda > 2 \times 10^{15} \, {\rm GeV}
\eeq
These are somewhat below the possible magnitudes 
$M, \Lambda = {\cal O}(10^{19})$ GeV~\footnote{The GRB shown in
Fig.~\ref{fig:GRB} has not yet been included in this analysis. With its
many
short-time structures and large redshift $z = 4.5$~\cite{fourpointfive},
its inclusion should strengthen these limits significantly.},
but not so far away!

Several exciting future steps seem possible. We expect the redshifts of
many GRBs to be measured shortly, using new early-warning satellites such
as HETE-II.  Also, more $\gamma$-ray telescopes sensitive to higher
energies are coming on-line.  In addition to ground-based telescopes such
as MILAGRO, satellites such as AMS and GLAST~\cite{GLAST} may make
important
contributions.  Another exciting possibility is to look for (the smearing
of) neutrino pulses from GRBs.  Some models predict that they might emit
observable $\nu$ fluxes at energies up to $10^{20}$ eV~\cite{Waxman}.  If
a GRB pulse
from $z \sim 1$ were seen at this energy, it would be sensitive to $M \sim
10^{26}$ GeV!  Conversely, if $M \sim 10^{19}$ GeV, such a neutrino pulse
would be spread over years of arrival times, and hence be
invisible~\cite{EMNV}! 

\subsection{Avoiding the GZK Cut-off?} 

Now let us return to the UHECR, and see how the possible modification of
Lorentz kinematics discussed in the previous Sections could be relevant to
their interpretation.  The point is that an energetic particle with
\beq
c_o^2 p^2 = E^2 (1 + E/M) + \cdots
\eeq
may not, when it strikes a low-energy photon, be able to create particles
via the reaction $p + \gamma_{CMB} \to \Delta \to n + \pi$ or $\gamma +
\gamma_{IRB} \to e^+ e^-$, because its energy is `too small' compared with
its momentum.  In this way the GZK cut-off might be avoided~\cite{GM,CG}.
However, before reaching this conclusion, it is necessary also to analyze
particle interactions and decays in a quantum-gravitational
framework~\cite{CEMN}: it is expected that energy would be conserved
only in a statistical sense~\cite{EHNS,Adler}.

\begin{figure}[htbp]
\begin{center}
\mbox{\epsfig{file=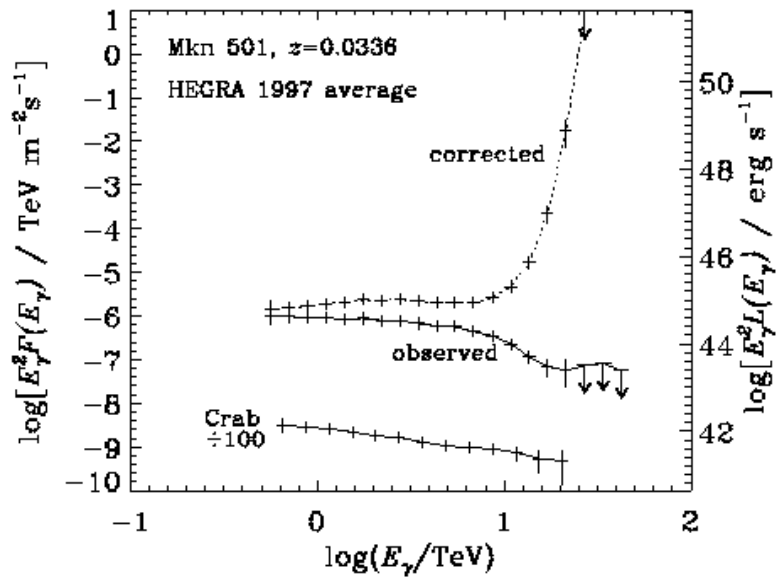,height=8cm}}
\end{center}
\caption[.]{\label{fig:Protheroe}\it
The signal observed by HEGRA~\cite{HEGRA} from Mkn~501, compared with
the corrected spectrum calculated using the expected absorption (see also 
Fig.~\ref{fig:cutoffs} by infra-red background photons~\cite{PM}.
}
\end{figure}

We have already discussed the GZK cut-off for cosmic-ray protons, and
noted its apparent absence.  It has recently been pointed out that the
corresponding cut-off on energetic astrophysical photons might also be
absent~\cite{PM}.  The story starts with the HEGRA report of TeV $\gamma$
rays from
Mkn 501~\cite{HEGRA}.  These should have been attenuated by absorption on
the infra-red
background.  The exercise has recently been performed of inverting this
attenuation, as seen in Fig.~\ref{fig:Protheroe}, to calculate what the
flux at the source would have to be in
order to produce the observed high-energy flux from Mkn 501~\cite{PM}. 
The
calculated source flux is remarkably high at energies $> 1$ TeV. 
Moreover, if the source were emitting so many energetic photons, then one
would expect it to emit comparably many neutrinos.  However, the AMANDA
neutrino telescope can rule out such a high neutrino flux~\cite{Barwick}. 

Does this mean that Lorentz kinematics is violated?  Surely not, since
there are many more prosaic interpretations of the data, starting with the
possibility that the HEGRA flux and/or energy calibration might need
adjustment, etc.  Nevertheless, this little mystery serves to remind us
that (at least some) quantum-gravitational speculations may not be very
far from experimental tests~\footnote{For example, as pointed out
by~\cite{AP}, there are approaches to quantum gravity that do not modify
the GZK cut-off.}. 

\subsection{Summary and Prospects}

In this lecture I have discussed two speculative fundamental physics ideas
that might be subject to tests using cosmic rays - the possibility that
UHECR might be due to the decays of ultraheavy metastable particles, and
the possibility that quantum gravity might modify the velocity of light. 
These provide two rival interpretations of the UHECR: the violation
of Lorentz kinematics in collisions might evade the GZK cut-off just as
well as relic decays.  

There are many other astrophysical sources suitable
for probing the proposed deviation from Lorentz kinematics, including GRBs
and AGNs.  Searches for timing delays for energetic $\gamma$'s have
already been used to limit possible deviations from the Lorentzian
momentum-energy relation for relativistic particles.  Also, there is the
puzzle of energetic $\gamma$ rays from Mkn 501 that could be resolved
simultaneously. Ultra-high-energy neutrinos from GRBs would be a great way
to probe deviations from Lorentz kinematics.

However, one should always remember that all conservative interpretations
should be tried and rejected before one embraces such a speculative
interpretation of cosmic-ray data.  The good news is that many new
experiments, such as Auger and EUSO for UHECR, and a new generation of
$\gamma$-ray experiments, will soon provide us with plenty of data to
compare with both conservative and radical hypothesis.  We may hope that
new fundamental physics will be revealed, but should nevertheless brace
ourselves for disappointment.


\begin{thebibliography}{99}


\bibitem{GZK}
K.~Greisen,
Phys.\ Rev.\ Lett.\  {16} (1966) 748;
G.~T.~Zatsepin and V.~A.~Kuzmin,
JETP Lett.\  {4} (1966) 78.

\bibitem{LS}
For a recent review, see: A. Letessier-Selvon, astro-ph/0006111.

\bibitem{noGZK}
N.~Hayashida {\it et al.},
Astrophys.\ J.\  {522} (1999) 225; and references therein.

\bibitem{BF}
G.~R.~Farrar and P.~L.~Biermann,
Phys.\ Rev.\ Lett.\  {81} (1998) 3579.

\bibitem{Gelmini}
J.~Ellis, T.~K.~Gaisser and G.~Steigman,
Nucl.\ Phys.\  {B177} (1981) 427;
J.~Ellis, G.~B.~Gelmini,
J.~L.~Lopez, D.~V.~Nanopoulos and S.~Sarkar,
Nucl.\ Phys.\  {B373} (1992) 399.

\bibitem{evade}
V.~Berezinsky, M.~Kachelriess and A.~Vilenkin,
Phys.\ Rev.\ Lett.\  {79} (1997) 4302.

\bibitem{ELN}
J.~Ellis, J.~L.~Lopez and D.~V.~Nanopoulos,
Phys.\ Lett.\  {B247} (1990) 257.

\bibitem{CKR}
D.~J.~Chung, E.~W.~Kolb and A.~Riotto,
Phys.\ Rev.\  {D59} (1999) 023501
and hep-ph/9810361;
D.~J.~Chung, E.~W.~Kolb, A.~Riotto and I.~I.~Tkachev,
Phys.\ Rev.\  {D62} (2000) 043508.

\bibitem{BEN}
K.~Benakli, J.~Ellis and D.~V.~Nanopoulos,
Phys.\ Rev.\  {D59} (1999) 047301.

\bibitem{Kach}
V.~Berezinsky and M.~Kachelriess,
Phys.\ Lett.\  {B434} (1998) 61 and
hep-ph/0009053.

\bibitem{BS}
M.~Birkel and S.~Sarkar,
Astropart.\ Phys.\  {9} (1998) 297.

\bibitem{Auger}
J.~J.~Beatty {\it et al.}, Auger Collaboration,
{\it The Pierre Auger Project Design Report},
FERMILAB-PUB-96-024.

\bibitem{Scarsi}
L.~Scarsi, talk at this meeting.

\bibitem{CLAF}
L.~Masperi and M.~Orsaria, talk at this meeting and
astro-ph/0008525.

\bibitem{Rovelli}
For a recent review and references, see: C.~Rovelli,
{\it Notes for a brief history of quantum gravity},
gr-qc/0006061.

\bibitem{Hawking}
S. Hawking, Commun.\ Math.\ Phys.\  {87} (1982) 395.

\bibitem{EHNS}
J.~Ellis, J.~S.~Hagelin, D.~V.~Nanopoulos and M.~Srednicki,
Nucl.\ Phys.\  {B241} (1984) 381.

\bibitem{AEMN}
G.~Amelino-Camelia, J.~Ellis, N.~E.~Mavromatos and D.~V.~Nanopoulos,
Int.\ J.\ Mod.\ Phys.\  {A12} (1997) 607.

\bibitem{Dmodel}
J.~Polchinski,
Phys.\ Rev.\ Lett.\  {75} (1995) 4724.

\bibitem{recoil}
J.~Ellis, N.~E.~Mavromatos and D.~V.~Nanopoulos,
Int.\ J.\ Mod.\ Phys.\  {A13} (1998) 1059.

\bibitem{EMN}
J.~Ellis, N.~E.~Mavromatos and D.~V.~Nanopoulos,
Gen.\ Rel.\ Grav.\ {31} (2000) 1257;
Phys.\ Rev.\  {D61} (2000) 027503.

\bibitem{EMNV}
J.~Ellis, N.~E.~Mavromatos, D.~V.~Nanopoulos and G.~Volkov,
Gen.\ Rel.\ Grav.\ {32} (2000) 1777.

\bibitem{EMNnull}
J.~Ellis, N.~E.~Mavromatos and D.~V.~Nanopoulos,
Phys.\ Rev.\ D61 (2000) 027503.

\bibitem{Mex}
J.~Alfaro, H.~A.~Morales-Tecotl and L.~F.~Urrutia,
Phys.\ Rev.\ Lett.\  {84} (2000) 2318.

\bibitem{Mitsou}
J.~Ellis, K.~Farakos, N.~E.~Mavromatos, V.~A.~Mitsou and D.~V.~Nanopoulos,
Astrophys.\ J.\  {535} (2000) 139.

\bibitem{horizon}
J.~Ellis, N.~E.~Mavromatos and D.~V.~Nanopoulos,
Phys.\ Rev.\  {D62} (2000) 084019.

\bibitem{higherD}
H.~Yu and L.~H.~Ford,
gr-qc/9907037 and
gr-qc/0004063;
A.~Campbell-Smith, J.~Ellis, N.~E.~Mavromatos and D.~V.~Nanopoulos,
Phys.\ Lett.\  {B466} (1999) 11.

\bibitem{thermal}
L.~J.~Garay, Phys.\ Rev.\ {D58} (1998) 124015.

\bibitem{loop}
C.~Rovelli and L.~Smolin, Phys.\ Rev.\ Lett.\ {61} (1988) 1155 and
Nucl.\ Phys.\ {B331} (1990) 80;
A.~Ashtekar, C.~Rovelli and L.~Smolin, Phys.\ Rev.\ Lett. {69} (1992)
237.


\bibitem{GP}
R.~Gambini and J.~Pullin,
Phys.\ Rev.\  {D59} (1999) 124021.

\bibitem{plasma1}
J.~Latorre, P. Pascual and R. Tarrach, Nucl.\ Phys.\ {B437} (1995) 60.

\bibitem{plasma2}
M. Thoma, hep-ph/0005282 and references therein.

\bibitem{AEMNS}
G.~Amelino-Camelia, J.~Ellis, N.~E.~Mavromatos, D.~V.~Nanopoulos and
S.~Sarkar,
Nature {393} (1998) 763.

\bibitem{fourpointfive}
M.I. Andersen {\it et al}. astro/ph0010322.

\bibitem{GRBs}
For a review of $\gamma$-ray bursters,
see
P.~Meszaros,
Nucl.\ Phys.\ Proc.\ Suppl.\  {80} (2000) 63.

\bibitem{EGRET}
For a review, see
R.~Mukherjee, asto-ph/9901222.

\bibitem{Milagrito}
R.~Atkins {\it et al.}, Milagro Collaboration,
astro-ph/0001111.

\bibitem{Schaefer}
B.~E.~Schaefer, astro-ph/9810479.

\bibitem{Biller}
S.~D.~Biller {\it et al.},
Phys.\ Rev.\ Lett.\  {83} (1999) 2108.

\bibitem{Kaaret}
P.~Kaaret, astro-ph/9903464.

\bibitem{GLAST}
J.~P.~Norris, J.~T.~Bonnell, G.~F.~Marani and J.~D.~Scargle,
astro-ph/9912136.

\bibitem{Waxman}
E. Waxman, hep-ph/0004102.

\bibitem{GM}
L.~Gonzalez-Mestres,
physics/0003080 and references therein.

\bibitem{CG}
S.~Coleman and S.~L.~Glashow,
Phys.\ Rev.\  {D59} (1999) 116008 and references therein.

\bibitem{CEMN}
A.~Campbell-Smith, J.~Ellis, N.~E.~Mavromatos and D.~V.~Nanopoulos,
in preparation.

\bibitem{Adler}
J.~Ellis, N.~E.~Mavromatos and D.~V.~Nanopoulos,
gr-qc/0007044 and references therein.

\bibitem{PM}
R.~J.~Protheroe and H.~Meyer,
astro-ph/0005349.

\bibitem{HEGRA}
F.A. Aharonian {\it et al.}, HEGRA Collaboration, Astron. Astrophys. 349
(1999) 11.

\bibitem{Barwick}
S. Barwick, for the AMANDA Collaboration, talk at Neutrino 2000.

\bibitem{AP}
G.~Amelino-Camelia and T.~Piran,
astro-ph/0006210 and astro-ph/0008107.

\end{thebibliography}
\end{document}